\documentclass[aps,prd]{revtex4}
\bibliography{apsrev}
\begin{document}

\title{ Quantum effects of a massive $3$-form coupled to a Dirac field.}

\author{Antonio Aurilia}
\email{aaurilia@csupomona.edu}
\affiliation{Department of Physics,
   California State Polytechnic University-Pomona}
\author{Euro Spallucci}
\email{spallucci@trieste.infn.it}
\affiliation{Dipartimento di Fisica Teorica, Universit\`a di Trieste
and INFN, Sezione di Trieste}
\date{\today}

\begin{abstract}
The computation of the quantum vacuum pressure must take into account
the contribution of zero-point oscillations of a rank-three gauge field
$A_{\mu\nu\rho}$. This result was established in a previous paper where we
calculated both the Casimir pressure within a region of vacuum simulating a
hadronic bag and the Wilson factor for the three-index
potential associated with the boundary of the bag. The resulting ``volume law''
satisfied by the Wilson loop is consistent with the basic confining
requirement that the static inter-quark potential increases with the distance
between two test charges. As a sequel to that paper, we consider here 
the coupling of $A_{\mu\nu\rho}$ to the generic
current of a matter field, later identified with the spin density current of a
Dirac field. In fact, one of the objectives of this paper is to investigate the
impact of the quantum fluctuations of $A_{\mu\nu\rho}$ on the effective
dynamics of the spinor field. The consistency of the field equations,
even at the classical level, requires the introduction of a mass term for
$A_{\mu\nu\rho}$. In this case, the Casimir vacuum pressure includes a
contribution that is explicitly dependent on the mass of $A_{\mu\nu\rho}$ and
leads us to conclude that the mass term plays the same role
  as the infrared cutoff needed to regularize the finite volume partition
functional previously calculated in the massless case. Remarkably, even in
the presence of a mass term, $A_{\mu\nu\rho}$ contains a mixture of massless
and massive spin-$0$ fields so that the resulting equation is still gauge
invariant. This is yet another peculiar, but physically relevant property of
$A_{\mu\nu\rho}$ since it is  reflected in the effective dynamics of the spinor
fields and confirms the confining property of $A_{\mu\nu\rho}$ already expected
from the earlier calculation of the Wilson loop.
   \end{abstract}

\pacs{42.50.Lc, 11.25-w}
\maketitle

\section{Introduction}

This the third in a series of articles devoted to an in-depth study
of the classical and quantum properties of a rank-three gauge form
$A_{\mu\nu\rho}$ in four spacetime dimensions
\cite{uno}, \cite{due}.\\
In a previous paper, hereafter referred to as
II \cite{due}, we have calculated the contribution to the
vacuum pressure, or Casimir energy density, due to the quantum
fluctuations of a \textit{ massless } antisymmetric tensor gauge field
$A_{\mu\nu\rho}$ with an associated field strength

\begin{equation}
F_{\lambda\mu\nu\rho}=\partial_{[\,\lambda}A_{\mu\nu\rho\,]} \label{uno}
\end{equation}

that represents, classically, a non-vanishing constant background field. \\
Our motivation for investigating the quantum properties of the
generalized Maxwell field
(\ref{uno}) was discussed in II. Here, we simply recall that the
$F$-field may be related
to two central issues in modern theoretical physics, namely, i) the
problem of dark matter/energy in the universe
via the cosmological constant  \cite{uno}, \cite{unob}, \cite{jap} 
and ii) the outstanding
problem of color confinement in $QCD$ via the "~bag constant~" \cite{tre},
\cite{treb}. Both constants, cosmological and hadronic,
can be effectively described by the $A_{\mu\nu\rho}$ field. \\
Apart from these physical considerations, the $F$-field is intriguing because
of its deceptive simplicity. Here we have essentially a "constant" disguised
as a gauge field, a situation all too familiar from the study of
"electrodynamics" in two  spacetime dimensions \cite{quattro}.
In actual fact, \textit{ a study of
$A_{\mu\nu\rho}$ is a study of the quantum vacuum} and some caution should be
exercised before dismissing it. Thus, a consistent quantum formulation of this
constant gauge field requires a modification of the ``~sum over histories
approach~'' in the sense that the path integral must take into account the
whole family of constant field configurations having  support only  over a
spacetime region with a finite volume $V$

\begin{equation}
\int [DF]\delta\left[\, \partial_\lambda\, F^{\lambda\mu\nu\rho}\,
\right]\exp\left(
-\frac{1}{2\times 4!}\int_Vd^4x
F^{\lambda\mu\nu\rho}F_{\lambda\mu\nu\rho}\,\right)
\propto
\int_{-\infty}^\infty df \exp\left(-\frac{f^2}{2}V\,\right)
   \end{equation}

where $f$ is an arbitrary constant  labeling each classical solution
of the free field equations

\begin{equation}
\partial_\lambda\, F^{\lambda\mu\nu\rho}=0\ .
\end{equation}

While the generalized Maxwell field (\ref{uno}) propagates no
physical quanta, it gives rise to a static effect, namely, a Casimir vacuum
pressure that is inversely proportional to the
volume of quantization $V$. This result was confirmed by an explicit
calculation of the vacuum
expectation value of the energy-momentum tensor associated with the $F$-field

\begin{equation}
\langle\, T_{\mu\nu}\,\rangle \vert_{g=\delta}=\Big\langle\,
\frac{1}{3!}F_{\mu\rho\sigma\tau}F_\nu^{\rho\sigma\tau}-\frac{1}{2\times
4!}\delta_{\mu\nu}
F^{\rho\sigma\tau\lambda}F_{\rho\sigma\tau\lambda}\,\Big\rangle=
\delta_{\mu\nu}\left(\, \rho_0 +\frac{1}{2V}\,\right)\ .\label{4}
\end{equation}

In the above expression, $\rho_0$ represents a classical contribution
to the vacuum energy and plays the role of an infrared cutoff that is
necessary in order to regularize the large volume behavior of the finite
volume partition function $Z(V)$ that was explicitly derived in II.\\
With hindsight, we recognize that the role of the infrared cutoff
$\rho_0$ may be played by  a mass term in the lagrangian for the $F$-field.
In the first order formalism this lagrangian reads

\begin{equation}
L_0= \frac{1}{2\times 4!}
F^{\rho\sigma\tau\lambda}F_{\rho\sigma\tau\lambda} -\frac{1}{4!}
F^{\lambda\mu\nu\rho}\partial_{[\,\lambda}A_{\mu\nu\rho\,]}-\frac{m^2_A}
{2\times 3!}A^{\mu\nu\rho}\, A_{\mu\nu\rho}\ .
\end{equation}

In spite of its unusual appearance, the above expression is a
Proca-type lagrangian that describes a massive spin-$0$ field \cite{cinque},
with some qualifications. Indeed, the field equations are

\begin{eqnarray}
&& F_{\lambda\mu\nu\rho}=\partial_{[\,\lambda}A_{\mu\nu\rho\,]}\ ,\label{6a}\\
&& \partial_\lambda\, F^{\lambda\mu\nu\rho} -m^2_A \,
A^{\mu\nu\rho}=0\ .\label{6b}
\end{eqnarray}

  From Eq.(\ref{6b}) follows the constraint equation

\begin{equation}
\partial_\lambda A^{\lambda\mu\nu}=0 \label{7}
\end{equation}

due to the antisymmetry of $F_{\lambda\mu\nu\rho}$.
Substituting equations (\ref{6a}) and (\ref{7}) into Eq.(\ref{6b}) we
arrive at the
Klein-Gordon equation

\begin{equation}
\left(\, -\partial^2 + m^2_A\,\right)  A^{\lambda\mu\nu}=0 \ .\label{8}
\end{equation}

On the other hand, the constraint (\ref{7}) implies that

\begin{equation}
A^{\lambda\mu\nu}=\frac{1}{m_A}\epsilon^{\lambda\mu\nu\rho}\partial_\rho\,\phi\
,\label{9}
\end{equation}

so that the scalar field $\phi$ satisfies the equation

\begin{equation}
\left(\, -\partial^2 +
m^2_A\,\right)\left(\,-\partial^2\,\right)\,\phi=0 \ .\label{9b}
\end{equation}
Several observations are in order here. First, it should be noted
that in the massless case
the Maxwell field strength possesses no physical degrees of freedom
in view of its invariance
under the tensor gauge transformation

\begin{equation}
\delta A_{\mu\nu\rho}=\partial_{[\,\mu}\lambda_{\nu\rho\,]}\ . \label{9c}
\end{equation}
In this connection, it is also worth observing that the Proca
equation for $A_{\mu\nu\rho}$
leads to a gauge invariant equation in spite of the presence of a
mass term. This is implicit
in the form of Eq.(\ref{9b}) due to the presence of the extra
d'Alembertian. Alternatively, in
terms of the dual field

\begin{equation}
{}^\ast A^\mu\equiv \frac{1}{3!}\epsilon^{\mu\nu\rho\sigma}\, A_{
\nu\rho\sigma}
=\frac{1}{ m_A}\,\partial^\mu\,\phi\ ,
\label{10}
\end{equation}

it follows from Eq.(\ref{8}) that

\begin{equation}
\left(\, -\partial^2 +m^2_A\,\right)\partial_\mu{}^\ast A^\mu=0
\end{equation}

which is invariant under the dual gauge transformation

\begin{equation}
\delta{}^\ast A^\mu= \frac{1}{3!}\epsilon^{\mu\nu\rho\sigma}
\partial_{[\,\nu}\lambda_{\rho\sigma\,]}\ .
\end{equation}

The advantage of using the first order formalism in the path integral
method as a means of
getting around the lengthy Fadeev-Popov procedure \cite{sei}  necessary to
eliminate those spurious degrees of freedom in the massless case was discussed
in II. We note, incidentally, that the presence of the extra D' Alambertian in
the massive case may be construed as further evidence
of the confining properties of the $A_{\mu\nu\rho}$-field already expected from
the earlier calculation of the Wilson loop \cite{due}.\\
A general solution of the field equations, in the first order formulation, can
be obtained after combining Eq.(\ref{6b}) and Eq.(\ref{9}) into the single
equation

\begin{equation}
\partial_\lambda\, F^{\lambda\mu\nu\rho}
=m_A\, \epsilon^{\mu\nu\rho\sigma}\partial_\sigma\,\phi \ .
\label{13}
\end{equation}
In Eq.(\ref{13}) the mass term plays the role of a source term for
the Maxwell field.
Accordingly, a formal expression for $F^{\lambda\mu\nu\rho}$ is as follows
\begin{equation}
F^{\lambda\mu\nu\rho}=\epsilon^{\lambda\mu\nu\rho}\, f -
m_A\,\partial^{[\,\lambda}\,
\frac{1}{-\partial^2}\,\epsilon^{\mu\nu\rho\,]\tau}\partial_\tau\,\phi
\label{16}
\end{equation}
where $f$ represents an arbitrary (~constant~) solution of the
associated  homogeneous
equation. It is precisely this constant background field $f$ that
gives rise to the Casimir
effect that we have investigated in II and summarized at the
beginning of this introduction.\\
Against this background, the purpose of this paper is twofold. First,
we wish to
investigate the contribution to the vacuum pressure due to the
quantum fluctuations of the
massive three-index field $A_{\mu\nu\rho}$. Our second objective is
to investigate the
effective theory of a spinor field confined to a volume of
quantization under the vacuum
pressure generated by the zero-point oscillations of $A_{\mu\nu\rho}$.\\
As anticipated earlier, we find the same volume
dependence of the vacuum pressure as in the massless case, plus an
extra contribution that
depends explicitly on the mass of $A_{\mu\nu\rho}$ and plays the same role
   of the phenomenological parameter $\rho_0$ reported in Eq.(\ref{4}).
These results are
obtained in Sect.2.\\
   Section.3 is devoted to the calculation of
the finite volume effective action of a spinor field. Our motivation is that
the gauge invariance of the $A_{\mu\nu\rho}$ field in the massless
case requires that its
classical matter counterpart be given by the three-dimensional world
history of a relativistic
membrane. At the quantum level this coupling may be naturally
realized through the
spin-density current of a Dirac field.
Our overall approach is nonperturbative to the extent that we sum over all
possible configurations of $A_{\mu\nu\rho}$ in order to determine its
quantum effects on the
effective dynamics of the spinor field. Section.4 concludes the paper with a
summary of our results and an outlook of further research into the quantum
properties of the $A_{\mu\nu\rho}$ field.

\section{Vacuum fluctuations of the massive $A_{\mu\nu\rho}$ field.  }

We have seen in the Introduction that there is a discontinuity
between the massless and
massive case in the field equation for $A_{\mu\nu\rho}$. In the
massless case, there are no
radiative degrees of freedom and the field strength describes a
non-vanishing constant
background field within a finite volume. There are, however, quantum
vacuum fluctuations
associated even with a constant
gauge field and the corresponding vacuum energy/pressure is explicitly given by
Eq.(\ref{4}) that was derived in II. In the massive
case, on the other hand, there are
physical quanta of spin-$0$ associated with $A_{\mu\nu\rho}$. It
seems reasonable, therefore,
to expect an additional contribution to the Casimir energy density/pressure
related to the mass of $A_{\mu\nu\rho}$. That this is the case can be
seen by calculating the
finite volume partition functional

\begin{equation}
Z(V)=\int \left[\, DF\, \right]\left[\, DA\, \right]\, \delta\left[\,
\partial_\mu
A^{\mu\nu\rho}\, \right]\exp\left(-\int d^4x\left[
\frac{1}{2\times 4!}
F^{\rho\sigma\tau\lambda}F_{\rho\sigma\tau\lambda} -\frac{1}{4!}
F^{\lambda\mu\nu\rho}\partial_{[\,\lambda}A_{\mu\nu\rho\,]}-
\frac{m^2_A}{2\times 3!}
A^{\mu\nu\rho}\, A_{\mu\nu\rho}\,\right]\, \right)\ .\label{17}
\end{equation}

Note that the expression (\ref{17}) explicitly takes into account the
constraint equation (\ref{7}).\\
Let us start the computation of $Z(V)$ from the $A$-integration.
Taking into account Eq.(\ref{9}), which is equivalent to
the constraint (\ref{7}), we find

\begin{equation}
Z(V)=\int \left[\, DF\, \right]\left[\, D\phi\,
\right]\,\exp\left(-\int d^4x\left[
\frac{1}{2\times 4!}
F^{\rho\sigma\tau\lambda}F_{\rho\sigma\tau\lambda} +\frac{1}{3!}
\left(\,\partial_\lambda\, F^{\lambda\mu\nu\rho}\,\right)
\epsilon_{\mu\nu\rho\sigma}\partial^\sigma\phi-\frac{m^2_A}{2}\phi^2\,\right]
\, \right)\ .\label{18}
\end{equation}

The form of Eq.(\ref{18}) suggests that it may be convenient to split the $F$-
field in the sum of
two parts  $\widehat F^{\lambda\mu\nu\rho}$ and  $\widetilde
F^{\lambda\mu\nu\rho}$
such that

\begin{eqnarray}
&& F^{\lambda\mu\nu\rho}=\widehat F^{\lambda\mu\nu\rho}+
\widetilde F^{\lambda\mu\nu\rho}\label{19}\\
&& \partial_\lambda\,\widehat F^{\lambda\mu\nu\rho}=0\ ,\label{20}\\
&&\partial_\lambda\,\widetilde F^{\lambda\mu\nu\rho}\ne 0\equiv h^{\mu\nu\rho}
\ ,
\qquad \partial_\mu\,h^{\mu\nu\rho}=0\ . \label{21}
\end{eqnarray}

Accordingly,
the functional integration over $F$ is factorized into an integration
over divergence free paths times an integration over remaining
histories parametrized by $h^{\mu\nu\rho}$. The latter integration is
restricted
to the subspace of divergence free $h$-configurations by the consistency
condition (\ref{21})

\begin{equation}
\left[\, DF\, \right]=\left[\, D\widehat F\, \right]\,\delta\left[\,
\partial_\lambda\,\widehat F^{\lambda\mu\nu\rho}\,\right]\,
\left[\, D\widetilde F\, \right]\,\left[\, Dh\, \right]\,   \delta\left[\,
\partial_\lambda\,\widetilde F^{\lambda\mu\nu\rho}-h^{\mu\nu\rho}\,\right]
\delta\left[\, \partial_\mu\,
h^{\mu\nu\rho}\,\right]\ .\label{23}
\end{equation}
Furthermore, Eq.(\ref{20}) implies that the corresponding functional measure
in (\ref{23}) must involve  only constant field configurations, namely,
   arbitrary constant solutions $f$ of Eq.(\ref{20}) with dimensions $\left[\,
f\,\right]=\mathrm{mass}^2$,

\begin{equation}
  \left[\, D\widehat F\, \right]\,   \delta\left[\,
\partial_\lambda\,\widehat F^{\lambda\mu\nu\rho}\,\right]\propto df
\left[\, D\widehat F\, \right]\,  \delta\left[\,\widehat F^{\lambda\mu\nu\rho}
-f\,\epsilon^{\lambda\mu\nu\rho}\,
\right] \ .
\end{equation}
Therefore, integrating over  $\widehat F$ means summing over all
constant $\widehat F$-field configurations that satisfy the
constraint (\ref{20}).
It seems worth observing, on the technical side, that this procedure
of integration is
identical to that introduced by the authors in the case of a free
point particle in which the integration over the momentum trajectories
is constrained to satisfy the classical equations of motion
$\dot{q^i}(t)=0$ \cite{prop}. In both cases,
the functional integration is reduced to an ordinary integration,
namely, $\left[\,Dq\,\right]\delta\left[\,\dot q^i(t)\,\right]
\longrightarrow d^3q/(2\pi)^3$.
However, by extracting an ordinary differential $df$ out of the dimensionless
functional measure $\left[\,
DF\, \right]$ we have to rescale the sum against
an arbitrarily fixed
standard measure $m_F^2$  with dimensions of a mass squared.\\
   Taking all of the above into account, the partition functional becomes

\begin{eqnarray}
&&
Z(V)=\frac{1}{m^2_F}\int_{-\infty}^{\infty}df\exp\left(-\frac{f^2}{2}V\,\right)
\int
\left[\, D\phi\, \right]\,\left[\, Dh^{\mu\nu\rho}\, \right]\,
\delta\left[\, \partial_\mu h^{\mu\nu\rho}\,
\right]\exp\left(-S\left[\,h^{\mu\nu\rho}\
,\phi\,\right]\,\right)\label{26}\\
&&S\left[\, h^{\mu\nu\rho}\ ,\phi\,\right]=\int d^4x\left[\,
-\frac{1}{2\times 3!}
h^{\mu\nu\rho}\frac{1}{-\partial^2} \, h_{\mu\nu\rho} +\frac{1}{4!}\phi \,
\epsilon_{\lambda\mu\nu\rho}\partial^{[\,
\lambda}h^{\mu\nu\rho\,]}-\frac{m^2_A}{2}\left(\,\partial\phi\,\right)^2\,
\right]\ .\label{27}
\end{eqnarray}

Since $h^{\mu\nu\rho}$ is divergenceless by definition, as seen in
Eq.(\ref{21}), our next step is to
rewrite it as the Hodge dual of a longitudinal four-vector:

\begin{equation}
h^{\mu\nu\rho}= \epsilon^{\mu\nu\rho\sigma}\,\partial_\sigma\xi\ .
\end{equation}

Then, integrating over $f$ and $\xi$ gives
\begin{eqnarray}
Z(V)&&=\left(\, \frac{2\pi}{V m^4_F}\,\right)^{1/2}\int
\left[\, D\phi\, \right]\,\left[\, D\xi\, \right]\,
\exp\left(-\int d^4x\left[\,\frac{1}{2}\partial_\mu\xi\,
\frac{1}{-\partial^2}\,
\partial^\mu\xi + \frac{m^2_A}{2}\,\left(\, \partial^2 \xi\,\right)
\left(\,\partial\phi\,\right)^2\,\right]
\,\right)\nonumber\\
&& =\left(\,\frac{2\pi}{V m^4_F}\,\right)^{1/2}\int\left[\, D\phi\,
\right] \exp\left(-\int
d^4x\left[\,\frac{1}{2}\phi\left( -\partial^2 +m^2_A
\,\right)\left(-\partial^2\,\right)\phi
\,\right]\,\right)\ .\label{32}
\end{eqnarray}

Note that this round of calculations has led us to a type of scalar
field equation
in the expression
(\ref{32}) that is precisely of the form given in Eq.(\ref{9b})
discussed in the Introduction. Note, in particular, the
characteristic presence of the double d'Alembertian that we
have linked to the
presence of a gauge invariant mass term.
The remaining integration over $\phi$ is gaussian and leads to the
formal result

\begin{eqnarray}
Z(V)&&\equiv \left(\,\frac{2\pi}{V m^4_F}\,\right)^{1/2} \exp\left[\, -W(\,
V\,)\,\right]\nonumber\\
&&=\left(\,\frac{2\pi}{V m^4_F}\,\right)^{1/2}
\left[\,\mathrm{det}\left(\,-\partial^2\,\right)\,\left(\,-\partial^2
+m_A^2\,\right)\,\right]^{-1/2}\ .\label{33}
\end{eqnarray}

In order to calculate the above determinant in closed form, the usual
procedure is to perform
the calculation in the large volume limit, namely, $V  m^4_A >> 1$.
In this limit, the
spectrum
of the operator $-\partial^2 + m^2_A$ can be treated as a continuum
so that the sum over the
eigenvalues can be approximated by the following integral

\begin{eqnarray}
W(\, V\,)=&&\frac{1}{2}\, V\int\frac{d^4k}{(2\pi)^4}\left[\,\ln\left(\,
k^2 +m^2_A \,\right) +\ln k^2\,\right]\nonumber\\
=&&-\frac{1}{2}\, V\int\frac{d^4k}{(2\pi)^4}\int_0^\infty
\frac{d\tau}{\tau}\left[\,
\exp\left[\, -\tau\left(\, k^2 +m_A^2\,\right)\,\right]  +
\exp\left(\, -\tau\,
k^2\,\right) \,\right]\ .\label{41}
\end{eqnarray}

Let us consider first the mass dependent term. We will see at the end
of this calculation
that the finite part contributing to $W(\, V\,)$ is actually
proportional to $m_A^4$.
Anticipating this result we drop altogether the contribution coming
from $\ln k^2$ in
the above expression. The expression (\ref{41}) is divergent in the limit
$\tau\to 0$, i.e., in the
short-distance regime.
Therefore, it is convenient to use the proper time regularization procedure to
cast Eq.(\ref{41}) in the following form

\begin{equation}
W_\epsilon(\, V\,)
=-\frac{1}{2}\frac{1}{(2\pi)^2}\, V\, \mu^{-2\epsilon} \int_0^\infty
\frac{d\tau}{\tau^{3+\epsilon}}\exp\left(\, -\tau\, m_A^2\,\right)\ ,\label{43}
\end{equation}

where we have  introduced a mass parameter $\mu$ so that $W$ is
a-dimensional. By rescaling
the proper time $x\equiv \tau\,m_A^2$, we obtain

\begin{eqnarray}
W_\epsilon(\, V\,)=&&-\frac{1}{2}\frac{1}{(2\pi)^2}\,
V\,m^4_A\left(\, \frac{\mu^2}{m^2_A}
\,\right)^{-\epsilon} \int_0^\infty dx\, x^{-3-\epsilon}
\exp\left(\, -x\right)\nonumber\\
=&&-\frac{1}{2}\frac{1}{(2\pi)^2}\, V\,m^4_A\left(\, \frac{\mu^2}{m^2_A}
\,\right)^{-\epsilon}\,\Gamma\left(\, -2-\epsilon\,\right)
\end{eqnarray}

where
\begin{eqnarray}
\Gamma\left(\,
-2-\epsilon\,\right)&&=-\frac{1}{\epsilon(1+\epsilon)(2+\epsilon)}
\Gamma\left(\, 1+\epsilon\,\right)\label{44}\\
&&\approx -\frac{1}{2\epsilon}
\left[\,1-\left(\, \gamma
+\frac{3}{2}\,\right)\epsilon+\dots\,\right]\label{46}\\
\gamma &&\equiv-\Gamma^\prime(0)=\,\qquad\hbox{Euler  Constant}\
\end{eqnarray}

and the expression in Eq.(\ref{46}) refers to the limit $\epsilon\to
0$. From here we
isolate the pole singularity from the convergent part of the integral
as follows

\begin{eqnarray}
W_\epsilon(\, V\,)\equiv  && W_\infty + W_{Reg.}\\
W_\infty\equiv &&\frac{1}{4\epsilon}\frac{1}{(2\pi)^2}\, V\,m^4_A\\
   W_{Reg.}\equiv &&-\frac{1}{4}\frac{1}{(2\pi)^2}\, V\, m^4_A\left[\,
\ln\left(\, \frac{\mu^2}{m^2_A}\,\right) +  \gamma+\frac{3}{2}\,\right]\ .
\end{eqnarray}
Then, in the  minimal subtraction scheme we obtain:

\begin{equation}
\ln Z_{Reg.}=-\frac{1}{2}\ln\frac{Vm_A^4}{2\pi}
-\frac{1}{4}\frac{1}{(2\pi)^2}\, V\,m^4_A\left[\,
\ln\left(\, \frac{m_A^2}{\mu^2}\,\right) -
\gamma-\frac{3}{2}\,\right]\ .\label{50}
\end{equation}

With this result in hands we are finally able to calculate the
vacuum pressure as follows

\begin{equation}
p\equiv -{\partial\over\partial V}\ln
Z_{Reg.}=\frac{1}{2V}+\frac{1}{4}\frac{1}{(2\pi)^2}\,
  m^4_A\left[\,\ln\left(\, \frac{m_A^2}{\mu^2}\,\right) -
\gamma-\frac{3}{2}\,\right]
\ .\label{51}
\end{equation}

This result should be compared with equation (\ref{4}) in the
Introduction. Equation(\ref{4})
represents the outcome of our calculation in the massless case
discussed in II . The first
term in Eq.(\ref{51}) represents the Casimir contribution which
dominates for small volumes
and agrees with the previous result for the massless case. The second
term represents a
new contribution that dominates in the large volume limit and
corresponds to the term $\rho_0$
that appears in Eq.(\ref{4}). This comparison gives an explicit expression for
$\rho_0$ in terms of the $A_{\mu\nu\rho}$ mass $m_A$ and regulator mass $\mu$.

\section{The interacting case.}

In our previous paper, II, we have calculated the Wilson loop for the
three-index potential associated with a bag with a boundary represented by the
three-dimensional world history of a closed spherical membrane. From the Wilson
loop we derived the static potential between two opposite points on the surface
of the bag and found it proportional to the volume enclosed by
the surface. This "volume law" seems a natural extension of the area
law found in the more conventional case of quantum chromodynamic strings and
is consistent with the basic underlying idea of confinement that it requires
an infinite amount of energy to  separate the two points \cite{area}.\\
It is against this background computation that we wish to consider
the coupling of $A_{\mu\nu\rho}$ to a current $J^{\mu\nu\rho}$ that is later
identified with the spin-density  current of a Dirac field.\\
Mathematically, this new situation amounts to taking as a new action
\begin{equation}
S=\int d^4x\left[\, \frac{1}{2\times 4!}
F^{\rho\sigma\tau\lambda}F_{\rho\sigma\tau\lambda} -\frac{1}{4!}
F^{\lambda\mu\nu\rho}\partial_{[\,\lambda}A_{\mu\nu\rho\,]}-\frac{m^2_A}
{2\times 3!} A^{\mu\nu\rho}\,
A_{\mu\nu\rho}-\frac{\kappa}{3!}A_{\mu\nu\rho}J^{\mu\nu\rho}\,\right]
\label{3.1}
\end{equation}

from which we find the field equations

\begin{eqnarray}
&& F_{\lambda\mu\nu\rho}=\partial_{[\,\lambda}A_{\mu\nu\rho\,]}\ ,\label{3.2}\\
&& \partial_\lambda\, F^{\lambda\mu\nu\rho} -m^2_A \,
A^{\mu\nu\rho}=\kappa J^{\mu\nu\rho}
\ .\label{3.3}
\end{eqnarray}

Equation (\ref{3.3}) was studied in I where the introduction of a
mass term was required by the
consistency of the field equations. As a matter of fact, without a mass term
Eq.(\ref{3.3}) is inconsistent as a
consequence of the antisymmetry of $F^{\lambda\mu\nu\rho}$ unless the
current on the right side
is divergence free. Thus, coupling $A^{\mu\nu\rho}$ to a spinor field
requires, in general, the
presence of a mass term. From Eq.(\ref{3.3}) follows the new
constraint equation
\begin{equation}
\partial_\lambda A^{\lambda\mu\nu}=-\frac{\kappa}{m_A^2}
\partial_\mu J^{\mu\nu\rho}\equiv
   -\frac{\kappa}{m_A^2}\,j^{\nu\rho}                 \label{3.4}
\end{equation}
   where we have defined $j^{\nu\rho}$ as the divergence of
$J^{\mu\nu\rho}$. A formal solution of
the constraint
Eq.(\ref{3.4}) is
\begin{equation}
A^{\lambda\mu\nu}=\frac{1}{m_A}\epsilon^{\lambda\mu\nu\rho}\partial_\rho\,\phi
-\frac{\kappa}{m_A^2}\,\partial^{[\,\lambda}\, \frac{1}{\partial^2}
\,  j^{\mu\nu\,]}
\ .\label{3.5}
\end{equation}
Substituting Eq.(\ref{3.5}) into Eq.(\ref{3.3}), we arrive at the
following field equation
\begin{equation}
\partial_\lambda\, F^{\lambda\mu\nu\rho}-m_A\,
\epsilon^{\mu\nu\rho\sigma}\partial_\sigma\,\phi=\kappa\left(\,
J^{\mu\nu\rho}-\partial^{[\,\mu}\, \frac{1}{\partial^2}\,j^{\nu\rho\,]}
\,\right) \ . \label{3.6}
\end{equation}
Therefore, in analogy to the free field case, Eq.(\ref{16}), a
generic solution of Eq.(\ref{3.6}) can be
expressed as follows

\begin{eqnarray}
F^{\lambda\mu\nu\rho}&&=\epsilon^{\lambda\mu\nu\rho}\, f +
m_A\,\partial^{[\, \lambda}
\, \frac{1}{\partial^2}
\,\epsilon^{\mu\nu\rho\,]\tau}\partial_\tau\,\phi +\kappa\,
\left(\, J^{\mu\nu\rho}-
\partial^{[\,\mu}\, \frac{1}{\partial^2} \, j^{\nu\rho\,]}\,\right)
\nonumber\\
&&\equiv\epsilon^{\lambda\mu\nu\rho}\, f + m_A\,\partial^{[\, \lambda}
\, \frac{1}{\partial^2}
\,\epsilon^{\mu\nu\rho\,]\tau}\partial_\tau\,\phi +\kappa\,
\widehat J^{\mu\nu\rho}\ .\label{3.5b}
\end{eqnarray}
Up until now, the nature of the three index current in the above
expressions has not been specified.
In our previous papers, both I and II, $J^{\mu\nu\rho}$ was
identified with the classical current
associated with the world history of a relativistic membrane.
Presently, as a stepping stone
toward a forthcoming discussion of the quantum chromodynamic case,
it is instructive to identify $J^{\mu\nu\rho}$ with the spin-density
current of a Dirac
field. As a matter of fact, as anticipated in the Introduction,
\textit{ one of the main purposes of this
paper is to determine the form of the
effective action } $\Gamma\left[\, \bar\psi\ ,\psi\,\right]$ \textit{
for a Dirac field
under the influence of the quantum vacuum fluctuations of } 
$A_{\mu\nu\rho}$. \\
In the massless case, considered in II, this same objective was achieved
by an explicit calculation
of the Wilson factor defined by
\begin{eqnarray}
W\left[\, J\,\right]=\Big\langle\, \exp\left(-\frac{\kappa}{3!}\int d^4x
A_{\mu\nu\rho}\,
J^{\mu\nu\rho}\,\right)\, \Big\rangle &&=\frac{Z\left[\, J\,\right]}{Z\left[\,
0\,\right]}\nonumber\\
&&\equiv \exp\left(-\Gamma\left[\, J\,\right]\,\right) \label{3.6b}
\end{eqnarray}
where
\begin{equation}
J^{\mu\nu\rho}=\int_{\partial B}\delta^{4)}\left[\, x - Y\,\right]\,
dy^\mu\wedge dy^\nu\wedge
dy^\rho \label{3.7}
\end{equation}
represents the classical current of a relativistic test bubble.
Presently we wish to undertake
the same calculation in the full quantum case in which the current
(\ref{3.7}) is replaced
by the spin density current of a spinor field in the background
vacuum that we have discussed
in the previous section
\begin{equation}
J^{\mu\nu\rho}\longrightarrow \Sigma^{\mu\nu\rho}\equiv
\bar\psi\gamma^\mu\gamma^\nu\gamma^\rho\psi \ . \label{3.8}
\end{equation}
Mathematically, our task is to integrate the partition functional
\begin{eqnarray}
Z\left[\,\Sigma\ , V\,\right]=&&\int \left[\, DF\, \right]\left[\, DA\,
\right]\, \delta\left[\, m^2_A\, \partial_\mu\,
A^{\mu\nu\rho}-\kappa\,\partial_\mu\,\Sigma^{\mu\nu\rho}\,\right]\times
\nonumber\\
&&\exp\left(-\int d^4x\left[ \frac{1}{2\times 4!}
F^{\rho\sigma\tau\lambda}F_{\rho\sigma\tau\lambda} -\frac{1}{4!}
F^{\lambda\mu\nu\rho}\partial_{[\,\lambda}A_{\mu\nu\rho\,]}
-\frac{m^2_A}{2\times 3!}
A^{\mu\nu\rho}\, A_{\mu\nu\rho}-\frac{\kappa}{3!}\,
\Sigma^{\mu\nu\rho}\, A_{\mu\nu\rho}
\,\right]\, \right)\ .\label{3.9}
\end{eqnarray}
Before proceeding with the calculations, let us introduce the notation
\begin{eqnarray}
&&J^{5\,\mu}\equiv \frac{1}{3!}\epsilon^\mu{}_{\nu\rho\tau}\Sigma^{\mu\nu\rho}\
,\label{3.10a}\\
&&j^{\nu\rho}\equiv \partial_\mu\,\Sigma^{\mu\nu\rho}=
\frac{1}{2}\epsilon^{\mu\nu\rho\tau}
\partial_{[\,\mu}\,J^5{}{\tau\,]}\ ,\label{3.10b}\\
&&J^{5\,\mu}\equiv J^{5\,\mu}{}_T +J^{5\,\mu}{}_L\ ,\label{3.10c}\\
&&J^{5\,\mu}{}_T=\left(\,\delta^\mu{}_{\nu}-\partial^\mu\,
\frac{1}{\partial^2}\, \partial_\nu
\,\right) J^{5\,\nu}\label{3.10d}\\
&&J^{5\,\mu}{}_L =\partial^\mu\, \frac{1}{\partial^2}\, \partial_\nu\,
J^{5\,\nu}\ .\label{3.10e}
\end{eqnarray}
Once again, we begin the computation of $Z\left[\,\Sigma\ ,
V\,\right]$ with the
$A$-integration. The constraint (\ref{3.4}) is encoded in the
functional Dirac-delta in
Eq.(\ref{3.9}), and enables us to replace $A$ with the classical
solution (\ref{3.5}).
Thus we obtain
\begin{eqnarray}
Z\left[\,\Sigma\ , V\,\right]&&=
\int \left[\, DF\, \right]\left[\, D\phi\, \right]\,
\exp\left(-\int d^4x\left[\,
\frac{1}{2\times 4!}F^{\rho\sigma\tau\lambda}F_{\rho\sigma\tau\lambda}
\right.\right.\nonumber\\
&&+\frac{1}{3!} \left. \left.
\left(\,\frac{1}{m_A}\epsilon_{\mu\nu\rho\sigma}\partial^\sigma\phi +
\frac{\kappa}{ m_A^2} \partial_{[\,\mu}\, \frac{1}{-\partial^2}
\,j_{\nu\rho\,]}\,\right)\partial_\lambda\, F^{\lambda\mu\nu\rho}
  -\frac{m^2_A}{2\times 3!}
\left(\,\frac{1}{m_A}\epsilon_{\mu\nu\rho\sigma}\partial^\sigma\phi +
\frac{\kappa}{ m_A^2} \partial_{[\,\mu}\, \frac{1}{-\partial^2}
\,j_{\nu\rho\,]}\,\right)^2 \right.\right.\nonumber\\
&&-\frac{\kappa}{3!}\, \Sigma^{\mu\nu\rho}\,\left.\left.
\left(\,\frac{1}{m_A}\epsilon_{\mu\nu\rho\sigma}\partial^\sigma\phi +
\frac{\kappa}{ m_A^2} \partial_{[\,\mu}\, \frac{1}{-\partial^2}
\,j_{\nu\rho\,]}\,\right)
  \,\right]\, \right)\ .\label{3.10}
\end{eqnarray}

The integration over $F$ proceeds exactly as in the free case.
Incidentally, this is one
advantage of the first order formalism in which $F$ and $A$ are
treated as independent
variables. Therefore

\begin{eqnarray}
Z\left[\,\Sigma\ , V\,\right]&&=\left(\,\frac{2\pi}{m_F^4V}\,\right)^{1/2}
\int \left[\, D\phi\, \right]\,
\left[\, D\xi\, \right]	\exp\left(-\int d^4x\left[\,
\frac{1}{2}\partial_\mu\xi\, \frac{1}{-\partial^2} \partial^\mu\xi
\right. \right.\nonumber\\
&&+\frac{1}{3!} \left.\left.
\left(\,\frac{1}{m_A}\epsilon_{\mu\nu\rho\sigma}\partial^\sigma\phi +
\frac{\kappa}{ m_A^2} \partial_{[\,\mu}\, \frac{1}{-\partial^2}
\,j_{\nu\rho\,]}\,\right)
\epsilon^{\lambda\mu\nu\rho}\partial_\lambda\, \xi
-\frac{m^2_A}{2\times 3!}
\left(\,\frac{1}{m_A}\epsilon_{\mu\nu\rho\sigma}\partial^\sigma\phi +
\frac{\kappa}{ m_A^2} \partial_{[\,\mu}\, \frac{1}{-\partial^2}
\,j_{\nu\rho\,]}\,\right)^2\right.\right.\nonumber\\
&& -\frac{\kappa}{3!}\, \Sigma^{\mu\nu\rho}\,\left. \left.
\left(\,\frac{1}{m_A}\epsilon_{\mu\nu\rho\sigma}\partial^\sigma\phi +
\frac{\kappa}{ m_A^2} \partial_{[\,\mu}\, \frac{1}{-\partial^2}
\,j_{\nu\rho\,]}\,\right)\,\right]\, \right)\ .\label{3.11b}
\end{eqnarray}

Next, we proceed with the integration of the $\eta$-field noting that
it is decoupled from
$j^{\nu\rho}$ since

\begin{equation}
\int d^4x
\epsilon^{\lambda\mu\nu\rho}\left(\,\partial_\lambda\xi\,\right)
\partial_{[\,\mu}\, \frac{1}{-\partial^2} \,j_{\nu\rho\,]}=-\int
d^4x\epsilon^{\lambda\mu\nu\rho}\,\xi\,\partial_{[\,\lambda}\partial_{[\,\mu}\,
\frac{1}{-\partial^2} \,j_{\nu\rho\,]\,]}\equiv 0\ . \label{3.11}
\end{equation}

Collecting our results so far, we find

\begin{eqnarray}
Z\left[\,\Sigma\ , V\,\right]&&=\left(\,\frac{2\pi}{m_F^4V}\,\right)^{1/2}
\int \left[\, D\phi\,
\right]\exp\left( -\int d^4x\left[\,
\frac{1}{2m_A^2}\left(-\partial^2\,\right)\phi\,\left(-\partial^2\,\right)\phi
\right.\right.\nonumber\\
&&-\frac{m^2_A}{2\times 3!}\left.\left.
\left(\,\frac{1}{m_A}\epsilon_{\mu\nu\rho\sigma}\partial^\sigma\phi +
\frac{\kappa}{ m_A^2} \partial^{[\,\mu}\, \frac{1}{-\partial^2}
\,j^{\nu\rho\,]}\,\right)^2
-\frac{\kappa}{3!}\, \Sigma^{\mu\nu\rho}\,
\left(\,\frac{1}{m_A}\epsilon_{\mu\nu\rho\sigma}\partial^\sigma\phi +
\frac{\kappa}{ m_A^2} \partial_{[\,\mu}\, \frac{1}{-\partial^2}
\,j_{\nu\rho\,]}\,\right)\,\right]\, \right)\ .\label{3.12}
\end{eqnarray}

Before integrating over $\phi$, it is convenient to collect
all the terms explicitly depending on $\phi$ and rewrite $Z\left[\,\Sigma\ ,
V\,\right]$ as follows

\begin{eqnarray}
Z\left[\,\Sigma\ , V\,\right]&&=\left(\,\frac{2\pi}{m_F^4V}\,\right)^{1/2}
\exp\left( -\frac{\kappa^2}{2m_A^2}
\int d^4x j^{\nu\rho}\,\frac{1}{-\partial^2}\, j_{\nu\rho}\,\right)
\times\nonumber\\
&& \int \left[\, D\phi\,
\right]\exp\left( -\int d^4x\left[\,
\frac{1}{2m_A^2}\left(-\partial^2\,\right)\phi\,\left(-\partial^2\,\right)\phi
+\frac{\kappa}{m_A}\,\phi\,\partial_\mu\,J^{5\,\mu}\,\right]\,\right)\
.\label{3.13}
\end{eqnarray}

The final integration over $\phi$ now leads to the following expression
\begin{eqnarray}
Z\left[\,\Sigma\ , V\,\right]&&=\left(\,\frac{2\pi}{m_F^4V}\,\right)^{1/2}
\left[\,\left(-\partial^2\,\right)
\left(-\partial^2+m_A^2\,\right)\,\right]^{1/2}\times\nonumber\\
&&
\exp\left(
-\frac{\kappa^2}{2}\int d^4x\left[\,
j^{\nu\rho}\,\frac{1}{-\partial^2}\, j_{\nu\rho}-
\left(\,\partial_\mu\,J^{5\,\mu}\,\right)\frac{1}{\left(-\partial^2\,\right)
\left(-\partial^2+m_A^2\,\right) }\left(\,\partial_\nu\,J^{5\,\nu}\,\right)
\,\right]\,
\right)\nonumber\\
&&=\sqrt{\frac{2\pi}{m_F^4V}}\left[\,\left(-\partial^2\,\right)
\left(-\partial^2+m_A^2\,\right)\,\right]^{1/2}\times\nonumber\\
&&
\exp\left( -\frac{\kappa^2}{2}
\int d^4x\left[\, \partial^{[\,\mu}\,
j^{5\,\nu\,]}_T\,\frac{1}{-\partial^2}\,
\partial_{[\, \mu}\, j^5_{T\,\nu\,]}-
J^{5\,\nu}_L\frac{1}{   \left(-\partial^2\,\right)
\left(-\partial^2+m_A^2\,\right) }\,J^5_{L\nu}
\,\right]\,\right)
\ .\label{3.14}
\end{eqnarray}

  From here we extract our final result: according to the definition
(\ref{3.6b}) the dynamics of the fermion field is governed by the effective
action  induced by quantum  fluctuations of the $A$-field,

\begin{equation}
\Gamma\left[\,\bar\psi\ , \psi\,\right]=\frac{\kappa^2}{2}
\int d^4x\left[\, \partial^{[\,\mu}\,
j^{5\,\nu\,]}_T\,\frac{1}{-\partial^2}\,
\partial_{[\, \mu}\, j^5_{T\,\nu\,]}-
J^{5\,\nu}_L\frac{1}{   \left(-\partial^2\,\right)
\left(-\partial^2+m_A^2\,\right) }\,J^5_{L\nu}
\,\right]
\ .\label{3.15}
\end{equation}
A cursory inspection of the above expression indicates that the original
interaction of the $A_{\mu\nu\rho}$ field coupled to the spin-density current
corresponds to an effective four-fermion interaction where two distinct
components can be identified. One is a long-range interaction that involves
only the transverse part of the axial current, the other is a short-range
interaction that involves only the longitudinal part of the axial
current. This dynamical splitting seems noteworthy to the extent that the Green
function of the longitudinal component corresponds, once again, to the dipole
operator of the scalar field equation (\ref{9b}) derived in the Introduction.
To our mind, this suggests that the longitudinal component of the axial current
behaves as a \textit{spin-0 field} which we interpret as a collective
excitation, or bound state, generated by the underlying dynamics of the spinor
field. According to our line of reasoning, this effect, which one may
call \textit{dynamical bosonization,} should be traced back to the properties
of the quantum vacuum created by the $A_{\mu\nu\rho}$ field.

\section{Summary and outlook}
In this paper we have resumed
an in-depth study of the properties of a rank three, antisymmetric tensor gauge
field $A_{\mu\nu\rho}$. \\
The classical properties of this field, and its
counterpart in any number of spacetime dimensions, have been known for a long
time \cite{sette}. Unlike its electromagnetic counterpart in four dimensions,
$A_{\mu\nu\rho}$ does not radiate photons, or any other type of physical wave.
However, very much like its ``electromagnetic'' counterpart in two dimensions,
it simply represents a constant background field. Thus, in the free case and
in flat spacetime (~of infinite extension~) $A_{\mu\nu\rho}$ cannot be
distinguished from the classical ``~vacuum~''. However, dismissing this
field as physically irrelevant on this basis, would be too hasty. Indeed, the
form of the gauge transformation (\ref{9c}) dictates that $A_{\mu\nu\rho}$
couples to extended objects represented by the world history of relativistic
membranes. Furthermore, even in the absence of interaction but in the presence
of a mass term, $A_{\mu\nu\rho}$ ``comes alive'' in the sense that it 
represents
a spin-0 field that obeys a Proca-type equation of the kind (\ref{6a}),
(\ref{6b}) discussed in the Introduction. The presence of a gauge 
invariant mass
term was linked in I to the Stueckelberg mechanism for $A_{\mu\nu\rho}$
\cite{cinqueb} and shown to be related to the production of dark energy/matter
in the universe.
The strong formal analogy between this peculiar mechanism of mass generation
in four dimensions and a similar mechanism in two dimensions was also
emphasized in I.\\
\\
To our mind, these unique properties of the $A$--field  suggest a 
rather fitting
model, at least qualitatively, of a \textit{hadronic bag} in which
the role of the phenomenological ``bag constant'' is taken over by the infrared
cutoff that is necessary to regularize the partition functional that describes
the quantum dynamics of $A_{\mu\nu\rho}$. In support of this
interpretation, it was necessary to show that there exists a
\textit{confining potential} within the bag. As a matter of fact, an explicit
calculation of the Wilson loop associated with a relativistic test
bubble simulating the surface of the bag, shows that the static
potential between any pair of diametrically opposite points on the surface of
the bag is proportional to the enclosed volume. We interpret this result as a
natural extension of the ``area law'' for confinement that has long been
established for chromodynamic strings. This result, in turn, paves
the way to the introduction of fermion fields in the model and gives logical
continuity to the calculations undertaken in this
paper concerning the quantum properties of $A_{\mu\nu\rho}$.\\

This paper extends our previous results in two ways. First, we have
calculated the contribution to the Casimir pressure due to the zero point
oscillations of $A_{\mu\nu\rho}$ in the \textit{massive case}. We find that
the critical quantum effect due to the integration constant of the homogeneous
massless equation, is still in place leading to the characteristic
Casimir pressure that is inversely proportional to the volume of the bag. In
addition, there is now a contribution to the vacuum pressure, Eq.(\ref{51}),
that is directly related to the mass of the  $A$-field and plays the same role
as the phenomenological bag constant introduced as an infrared cutoff
in the partition functional for the massless case.\\

The second extension concerns the effect of the newly derived
vacuum pressure on the effective dynamics of fermion fields. This
amounts, in practice, to a recalculation of the Wilson loop extended to the
full quantum case in which the classical current
(\ref{3.7}) is replaced by the spin-density current (\ref{3.8}) of
the fermion field. The result is displayed in Eq. (\ref{3.15}) and is
noteworthy for the explicit presence of the quantum propagator
associated with the field equation (\ref{9b}) discussed in the
Introduction. \\

Just as in the case of the confining static potential associated
with a relativistic test bubble, the dipole
structure of the quantum propagator in Eq. (\ref{3.15}) in the
present case is symptomatic of the confining and screening properties of the
$A_{\mu\nu\rho}$ field at the quantum level.\\
We have repeatedly emphasized the stringent similarity between these
properties of the $A$-field in four dimensions and those of the
``electromagnetic'' field in two space-time dimensions. The
exact nature of this analogy and its possible impact on the problem
of color confinement in QCD will be discussed in a forthcoming publication.

\end{document}